\documentclass{iopjournal}
\usepackage{multirow}
\usepackage{amssymb}
\usepackage{graphicx}
\usepackage{tikz}
\usetikzlibrary{arrows.meta}
\usepackage{amsmath, amssymb}
\usepackage{silence}
\usepackage[compatibility=false]{caption}
\usepackage{subcaption}
\usepackage[T1]{fontenc}
\usepackage{lmodern}
\usetikzlibrary{arrows.meta, calc}

\begin{document}

\articletype{Paper} %	 e.g. Paper, Letter, Topical Review...

\title{Manifold Learning for Source Separation in Confusion-Limited Gravitational-Wave Data}

\author{Jericho Cain$^{1,*}$\orcid{0000-0003-4731-9142}}

\affil{$^1$Physics Department, Portland Community College, Portland, OR, USA}

\affil{$^*$Author to whom any correspondence should be addressed.}

\email{jericho.cain@gmail.com}

\keywords{gravitational waves, machine learning, manifold learning, continuous wavelet transform, signal detection, topology, smooth manifolds, LISA}

\begin{abstract}
The Laser Interferometer Space Antenna (LISA) will observe gravitational-waves in a regime that differs from what ground-based detectors handle. Instead of searching for rare signals buried in loud instrumental noise, LISA's main challenge is that its data stream contains millions of unresolved galactic binaries. These blend together into a confusion background, and the problem becomes distinguishing sources that genuinely stand out from that sea of signals. In this work we explore whether manifold-learning tools can help with that separation task using a controlled synthetic LISA-like dataset containing instrumental noise and simulated confusion backgrounds. We built a convolutional neural network (CNN) autoencoder trained solely on the confusion background. The model operates on scalograms produced using the continuous wavelet transform (CWT) and provides a reconstruction error, denoted $\epsilon(x)$, which measures how well the input scalogram lies on the learned background manifold. To incorporate geometric information from the latent space, we introduce an additional anomaly term, $\delta_{\perp}(\phi(x)$, defined as the off-manifold distance obtained from a local tangent-space estimate. The combined anomaly score therefore takes the form
\[
s(x) = \alpha \cdot \epsilon(x) + \beta \cdot \delta_{\perp}(\phi(x)).
\]
Performance is evaluated using the area under the receiver operating characteristic curve (ROC-AUC) and average precision (AP). A grid search over $\alpha$ and $\beta$ in the combined score revealed the best performance near $\alpha = 0.5$ and $\beta = 2.0$. This indicates that the latent-space geometry provides additional discriminative information beyond the reconstruction error alone. With this combination, the method reaches ROC-AUC $= 0.752$ and AP $= 0.810$; at the threshold that optimizes the F1 score, precision $= 0.81$ and recall $= 0.61$. This corresponds to roughly a $35\%$ improvement in ROC-AUC over using the autoencoder alone. The comparatively large coefficient on the manifold term ($\beta = 2.0$) suggests that the latent space captures meaningful geometric structure associated with the confusion background. Overall, these results indicate that manifold-learning techniques could provide a useful complementary tool in LISA data-analysis pipelines for identifying resolvable sources within a heavily confusion-limited dataset.
\end{abstract}

\section{Introduction}
\label{sec:introduction}

LISA will open a new window into gravitational-wave astronomy by observing the millihertz frequency band \cite{AmaroSeoane2017}. Ground-based detectors, while highly successful, operate in the tens to hundreds of hertz range. With ground-based observatories the primary challenge is detecting sparse, transient signals buried in high instrumental noise. LISA faces a fundamentally different problem: the confusion-limited regime. The LISA band will contain millions of unresolved galactic binaries which creates a stochastic background that dominates the data stream \cite{Nelemans2001,Cornish2017}. Hidden within this confusion noise are rare but scientifically valuable sources such as massive black hole binaries (MBHBs) and extreme mass ratio inspirals (EMRIs). This paper addresses one potential approach to this source separation problem: how to disentangle interesting signals from a sea of overlapping background sources.

Traditional approaches often rely on matched-filter techniques such as
the $\mathcal{F}$-statistic and related continuous-wave search methods \cite{Jaranowski1998},
as well as global Bayesian inference frameworks that attempt to jointly fit
large populations of sources \cite{Crowder2007}. These methods have proven powerful in many
gravitational-wave analyses, but they become computationally challenging in
the LISA confusion-limited regime. The vast parameter space of potential
source types, together with the presence of many overlapping signals,
makes large-scale template searches and global inference computationally
demanding.

Considerable effort has already been devoted to developing analysis methods for the
LISA confusion-limited regime. Early studies explored this problem through the
LISA Mock Data Challenges, which established benchmark datasets and stimulated
the development of a variety of search strategies for overlapping galactic binary
signals \cite{Arnaud2006}. More recent work within the LISA Data Challenge (LDC) program continues
to investigate scalable methods for extracting resolvable sources from the
confusion foreground \cite{Katz2022}.

Machine-learning methods have increasingly been applied to gravitational-wave
data analysis. Examples include glitch classification through the Gravity Spy
project \cite{Zevin2017}, time-frequency burst searches using Omega scans
\cite{Chatterji2004}, and neural-network approaches to continuous-wave
searches \cite{Dreissigacker2020}. More broadly, machine learning has been
explored for tasks such as glitch classification \cite{George2018},
signal denoising \cite{Wang2022}, and template-free gravitational-wave
detection \cite{Cain2026}. These studies demonstrate that data-driven
approaches can complement traditional analysis pipelines, particularly
when signal morphology is complex or the parameter space becomes large.
However, most existing machine-learning work has focused on the
signal-versus-noise detection problem characteristic of ground-based
detectors, rather than the source separation challenge inherent to
LISA's confusion-limited observations.

The manifold hypothesis posits that high-dimensional data often lies near a
low-dimensional smooth manifold. In the present context, the relevant manifold
is not a spacetime or signal-parameter manifold, but rather the geometric
structure traced by confusion-background scalograms in the learned
representation space. If the confusion background occupies a well-defined
low-dimensional structure in this space, then resolvable sources that deviate
from the background distribution should be identifiable through their geometric
properties rather than through reconstruction error alone. The continuous wavelet transform (CWT) preprocessing
helps preserve the time-frequency morphology of different source classes.
Massive black hole binaries, with their characteristic chirp signatures,
extreme mass ratio inspirals with their complex orbital dynamics, and galactic
binaries with their nearly monochromatic signals occupy distinct regions of the
time-frequency plane. An autoencoder trained on confusion background can then
learn a latent representation that captures this geometric structure.

To our knowledge, manifold-learning approaches have not yet been applied specifically to the problem of identifying resolvable sources within LISA's confusion-limited gravitational-wave data. We also could not find a machine learning approach that has been developed specifically for identifying resolvable sources in LISA's confusion-limited regime. While extensive literature exists on manifold learning in general \cite{Tenenbaum2000,Roweis2000}, and the autoencoder manifold hypothesis has been explored in various contexts \cite{Bengio2013}, these techniques have not been adapted to the unique challenges of gravitational-wave source separation. Similarly, while LISA science requirements and confusion noise modeling have been extensively studied \cite{AmaroSeoane2017}, machine learning approaches for source separation in this context remain unexplored.

Here we present the first application of manifold learning for signal separation in LISA-like gravitational-wave data. Inspired by the CWT-Long Short-Term Memory (LSTM) autoencoder architecture previously developed for LIGO detection \cite{Cain2026}, we adapt the approach for the LISA source separation problem. Specifically, we use a CWT-convolutional neural network (CNN) autoencoder and augment reconstruction error with geometric priors derived from the latent space manifold. Our approach trains an autoencoder on synthetic LISA-like data containing instrumental noise and confusion background, then constructs a k-nearest neighbor manifold in the latent space to quantify geometric deviations. We evaluate the combined anomaly score, $\alpha \cdot \epsilon(x) + \beta \cdot \delta_{\perp}(\phi(x))$, on test sets containing resolvable sources embedded in confusion noise, demonstrating that manifold geometry provides significant discriminative power beyond reconstruction error alone. Here, $\epsilon(x)$ measures how well the input scalogram lies on the learned background manifold and $\delta_{\perp}(\phi(x))$ is defined as the off-manifold distance obtained from a local tangent-space estimate. Our results show that the optimal configuration ($\alpha = 0.5$, $\beta = 2.0$) achieves receiver operating characteristic area under the curve (ROC-AUC) = 0.752 and average precision (AP) = 0.810; at the threshold that optimizes F1 score, precision = 0.81 and recall = 0.61. This represents a 35\% relative improvement in ROC-AUC over autoencoder-only detection. The fact that $\beta = 2.0$, compared with \(\alpha = 0.5\), indicates that manifold geometry contributes four times more weight than reconstruction error in the optimal combined score, revealing strong geometric structure in the latent space that can be leveraged for source separation.

This paper is organized as follows. Section~\ref{sec:data} describes our synthetic LISA-like data generation, including instrumental noise modeling, waveform generation for different source types, and confusion noise construction. Section~\ref{sec:preprocessing} details the continuous wavelet transform preprocessing adapted for LISA's frequency range. Section~\ref{sec:autoencoder} provides background on autoencoders in the context of smooth manifolds and topology. Section~\ref{sec:manifold} also presents our manifold learning framework, including tangent space estimation and the geometric decomposition of latent vectors. Section~\ref{sec:experimental} presents our experimental setup, including data distributions, model architecture, and hyperparameters. Section~\ref{sec:results} reports our results and interpretation. Section~\ref{sec:conclusion} concludes with a discussion of implications and future directions.

\section{Synthetic Data Generation}
\label{sec:data}

We generate synthetic LISA data following established models for instrumental noise, gravitational waveforms, and galactic confusion background. Our implementation follows the LISA Science Requirements Document and published sensitivity curves \cite{Cornish2017}.

\subsection*{Instrumental Noise}
\medskip

LISA's instrumental noise is characterized by a power spectral density (PSD) $S_n(f)$ with two dominant components. At low frequencies, acceleration noise from residual forces on the test masses dominates, while at high frequencies, optical metrology system (OMS) noise from laser phase measurement limits sensitivity. The strain noise PSD is given by:

\begin{equation}
S_n(f) = \frac{10}{3L^2} \left[ S_{\text{OMS}} + \frac{(3 + \cos(2f/f_*)) S_{\text{acc}}}{(2\pi f)^2} \right]
\end{equation}

where $L = 2.5 \times 10^9$ m is the arm length, $f_* = 19.09$ mHz is the transfer frequency, and the component noise PSDs are:

\begin{align}
S_{\text{acc}}(f) &= (3 \times 10^{-15} \text{ m/s}^2/\sqrt{\text{Hz}})^2 \left(1 + \left(\frac{0.4 \text{ mHz}}{f}\right)^2\right) \left(1 + \left(\frac{f}{8 \text{ mHz}}\right)^4\right) \\
S_{\text{OMS}} &= (15 \times 10^{-12} \text{ m}/\sqrt{\text{Hz}})^2
\end{align}

Time-domain noise realizations are generated by coloring white Gaussian noise with this PSD using the inverse Fourier transform, ensuring the noise statistics match the expected LISA sensitivity curve.

\subsection*{Gravitational Waveform Generation}

\medskip
We generate simplified analytical waveforms for three source types that span LISA's science goals. For massive black hole binaries (MBHBs), we use a post-Newtonian approximation with frequency evolution $f(t) = f_0 (1 - t/\tau)^{-3/8}$ and amplitude scaling $h_0 \propto (G M_{\text{chirp}}/c^2)^{5/3} (\pi f/c)^{2/3} / d_L$, where $M_{\text{chirp}}$ is the chirp mass and $d_L$ is the luminosity distance. The waveform includes both plus and cross polarizations with inclination-dependent amplitudes.

For extreme mass ratio inspirals (EMRIs), we generate quasi-periodic waveforms with multiple harmonics, capturing the complex orbital dynamics around a central massive black hole. The waveform includes frequency evolution from gravitational radiation reaction and orbital precession effects.

Galactic binaries are modeled as nearly monochromatic sources with slow frequency evolution, $h(t) = A \cos(2\pi f_{\text{gw}} t + \pi \dot{f} t^2 + \phi_0)$, where the amplitude $A$ depends on the binary masses, distance, and orbital inclination.

All waveforms are generated at a sampling rate of 1 Hz for durations of 3600 seconds (1 hour segments), matching LISA's expected data stream characteristics.

The waveform models used here intentionally omit several effects that are
important in fully realistic LISA analyses. In particular, we do not include
Doppler modulation and Roemer-delay effects arising from the orbital motion
of the detector, nor the associated dependence on sky location. These
simplifications allow us to construct a controlled proof-of-concept dataset
focused on the geometric properties of the confusion background. Incorporating
sky-location dependence, Doppler modulation, and longer coherent observations
would be natural extensions for future work aimed at more realistic LISA
simulations.

\subsection*{Confusion Noise}

\medskip
The galactic confusion background is modeled as the incoherent sum of many unresolved galactic binaries. Each confusion source is a weak galactic binary with signal-to-noise ratio (SNR) in the range $[0.1, 2.0]$, below the individual detection threshold. In our simulations we generate $N_{\text{conf}} = 1000$ such sources per segment, each with randomly sampled frequency, amplitude, and phase parameters. This simplified population serves as a tractable approximation to the much larger astrophysical population expected in the LISA band, which is predicted to contain millions of unresolved galactic binaries. The total confusion signal is:

\begin{equation}
h_{\text{confusion}}(t) = \sum_{i=1}^{N_{\text{conf}}} h_{\text{GB},i}(t)
\end{equation}

This creates a stochastic background that mimics the expected LISA confusion noise from millions of unresolved galactic binaries, while remaining computationally tractable for our training datasets.

\subsection*{Dataset Construction}

\medskip
Our training set consists of 5000 background segments, where each segment denotes a
3600\,s (one hour) time window containing instrumental noise and confusion
background. These background segments represent the typical LISA data stream
without additional resolvable sources. The test set contains 200 background
segments and 400 signal segments. Each signal segment contains the same
background realization together with a single injected resolvable source
(MBHB, EMRI, or galactic binary). Resolvable sources have SNR in the range
$[10, 50]$, ensuring that they are detectable above the confusion background
within this controlled simulation setup. The signal type distribution is
50\% MBHB, 30\% EMRI, and 20\% galactic binary, reflecting the relative
scientific importance and expected rates of these sources.

All signals are scaled to achieve the target SNR using the matched-filter definition:

\begin{equation}
\text{SNR}^2 = 4 \int_0^\infty \frac{|\tilde{h}(f)|^2}{S_n(f)} df
\end{equation}

where $\tilde{h}(f)$ is the Fourier transform of the signal and $S_n(f)$ is the noise PSD. This ensures that resolvable sources are consistently detectable above the background, while confusion sources remain below the detection threshold.

\section{Continuous Wavelet Transform Preprocessing}
\label{sec:preprocessing}

The continuous wavelet transform (CWT) provides a natural time-frequency representation for LISA gravitational-wave data that aligns with both the physical characteristics of the sources and the geometric structure required for manifold learning. Unlike Fourier-based methods that assume stationarity, CWT captures the non-stationary, evolving nature of gravitational-wave signals while preserving the geometric relationships between different source types.

\subsection*{Motivation for CWT}

\medskip
The choice of CWT for LISA data analysis is motivated by several factors. The geometry of different source types is directly visible in the time-frequency plane. Galactic binaries appear as nearly stationary horizontal lines, extreme mass ratio inspirals follow slowly sweeping, highly curved paths, and massive black hole binaries produce fast chirps that rapidly increase in frequency. These patterns are curves embedded in a two-dimensional time-frequency surface. 

Additionally, the confusion background manifests as overlapping time-frequency patterns from many unresolved sources. Manifold learning and clustering can operate on whole scalograms or local patches in time-frequency space to separate which features belong to which underlying source manifold. This geometric perspective is more intuitive than working directly in the time domain where overlapping signals are difficult to visualize and separate.

LISA signals are also extremely long in the time domain. A one-hour segment sampled at 1 Hz contains 3600 samples. Longer observations would be computationally expensive for direct time-domain processing. CWT provides time localization while offering some compression and enabling more efficient windowing and tiling strategies. The visual aspect of scalograms also helps when debugging and with interpretation of model behavior as failures can be directly observed in the time-frequency plane.

\subsection*{Implementation}

\medskip
We adapt the CWT implementation from our previous LIGO work \cite{Cain2026} to LISA's frequency range and time scales. The key changes are the frequency range (0.1 mHz to 100 mHz for LISA versus 20-512 Hz for LIGO), sampling rate (1 Hz versus 4096 Hz), and signal duration (3600 seconds versus 32 seconds). We use the Morlet wavelet because it provides a good time-frequency localization for chirping signals.

We first apply a fourth-order Butterworth high-pass filter at $f_{\text{min}} = 0.1$ mHz to remove low-frequency drifts. Then we whiten the signal by subtracting the global mean and dividing by the global standard deviation. These statistics are computed from the raw time-domain training background data. This global normalization prevents batch effects that can arise when normalizing each segment independently.

Next, we compute the CWT using logarithmically spaced frequency scales. The number of scales $n_{\text{scales}} = 140$ is chosen to match approximately 14 scales per octave. This provides adequate frequency resolution across LISA's frequency range. The scales are computed as $s = f_s / f$ where $f_s$ is the sampling rate and $f$ ranges logarithmically from $f_{\text{min}}$ to $f_{\text{max}}$. The resulting CWT therefore produces 140 pseudo-frequency samples determined
by the chosen wavelet scales. For use in the convolutional autoencoder, the
scalogram is subsequently resized to a fixed frequency resolution of 100 bins
using interpolation. This step standardizes the input dimensions across all
segments while preserving the overall time-frequency structure present in the
original CWT representation.

We then take the magnitude of the complex CWT coefficients to obtain the scalogram. We then use a logarithmic transform, $\log_{10}(|\text{CWT}| + \epsilon)$ where $\epsilon = 10^{-10}$ prevents numerical issues. This compresses the dynamic range of the scalogram. Finally, we normalize each scalogram segment to zero mean and unit variance after the log transform.

The scalogram is resized to target dimensions of $100 \times 3600$ (frequency scales $\times$ time samples) using bilinear interpolation. The target height of 100 frequency bins provides sufficient resolution to capture the geometric structure of different source types. The width of 3600 samples corresponds to one hour of data at 1 Hz sampling. This fixed-size representation is required for batch processing in the autoencoder architecture.

The global normalization statistics are computed from a sample of training background segments before CWT preprocessing. This ensures that all segments in both the training and test data are normalized consistently. This prevents the model from learning spurious correlations related to segment-to-segment variations in noise level.

\section{Autoencoder Architecture}
\label{sec:autoencoder}

We employ a convolutional autoencoder architecture that learns a low-dimensional representation of LISA-like CWT scalograms. The autoencoder serves a dual purpose: it provides reconstruction error as an anomaly score, and it learns a latent space where we can analyze the geometric structure of the data manifold. This geometric perspective, which we formalize in Section~\ref{sec:manifold}, motivates our approach to source separation.

\subsection*{Geometric Interpretation}

\medskip
From a differential geometry perspective, the autoencoder learns a chart from the data space to a low-dimensional latent space. Let $\mathcal{X} \subset \mathbb{R}^n$ be the space of CWT scalograms, where $n = 100 \times 3600 = 360{,}000$ for our simulated LISA data. According to the  manifold hypothesis, the trainingdata(confusion background) lies near a smooth manifold $\mathcal{M} \subset \mathcal{X}$ of much lower intrinsic dimension $d \ll n$.

The encoder $\phi: \mathcal{X} \to \mathcal{Z}$ maps scalograms to a $d$-dimensional latent space $\mathcal{Z} = \mathbb{R}^d$, where $d = 32$ in our implementation. For data $x \in \mathcal{M}$ near the manifold, the encoder approximates a chart, providing local coordinates $\phi(x) = z \in \mathcal{Z}$. The decoder $\psi: \mathcal{Z} \to \mathcal{X}$ approximates the inverse chart, mapping latent coordinates back to the data space. The composite map $\psi \circ \phi: \mathcal{X} \to \mathcal{X}$ attempts to reconstruct points on the manifold, with reconstruction error $\epsilon(x) = \|x - \psi(\phi(x))\|^2$ measuring how well the point lies on the learned manifold. From a statistical perspective, the reconstruction error $\epsilon(x)$ acts as a test statistic for detecting deviations from the learned confusion-background manifold. Segments that lie close to the background manifold produce small reconstruction errors, while segments containing additional signal structure tend to yield larger values of this statistic. Standard detection metrics such as the ROC-AUC can then be used to evaluate the resulting trade-off between detection efficiency and false-alarm rate.

The use of convolutional layers is important for this approach. Convolutions preserve local structure in the time-frequency plane. This is so nearby regions in the scalogram are processed together. This locality means that the learned representation respects the smooth structure of the manifold meaning that small perturbations in the time-frequency domain correspond to small changes in the latent space which is important for the manifold to be well-behaved. Also, translation invariance in the convolutional layers means that the same time-frequency patterns are encoded similarly regardless of their position. This helps the model learn the actual geometry rather than position-dependent artifacts.

This geometric framework lends itself directly to anomaly detection: points that lie on the manifold (confusion background) have low reconstruction error while points that lie some distance from the manifold (resolvable sources) should have high reconstruction error. However, reconstruction error alone may not capture all geometric information, which motivates our augmentation with explicit manifold geometry in Section~\ref{sec:manifold}. 
\subsection*{Architecture}

\begin{figure}[htbp]
    \centering
    \includegraphics[width=0.95\textwidth]{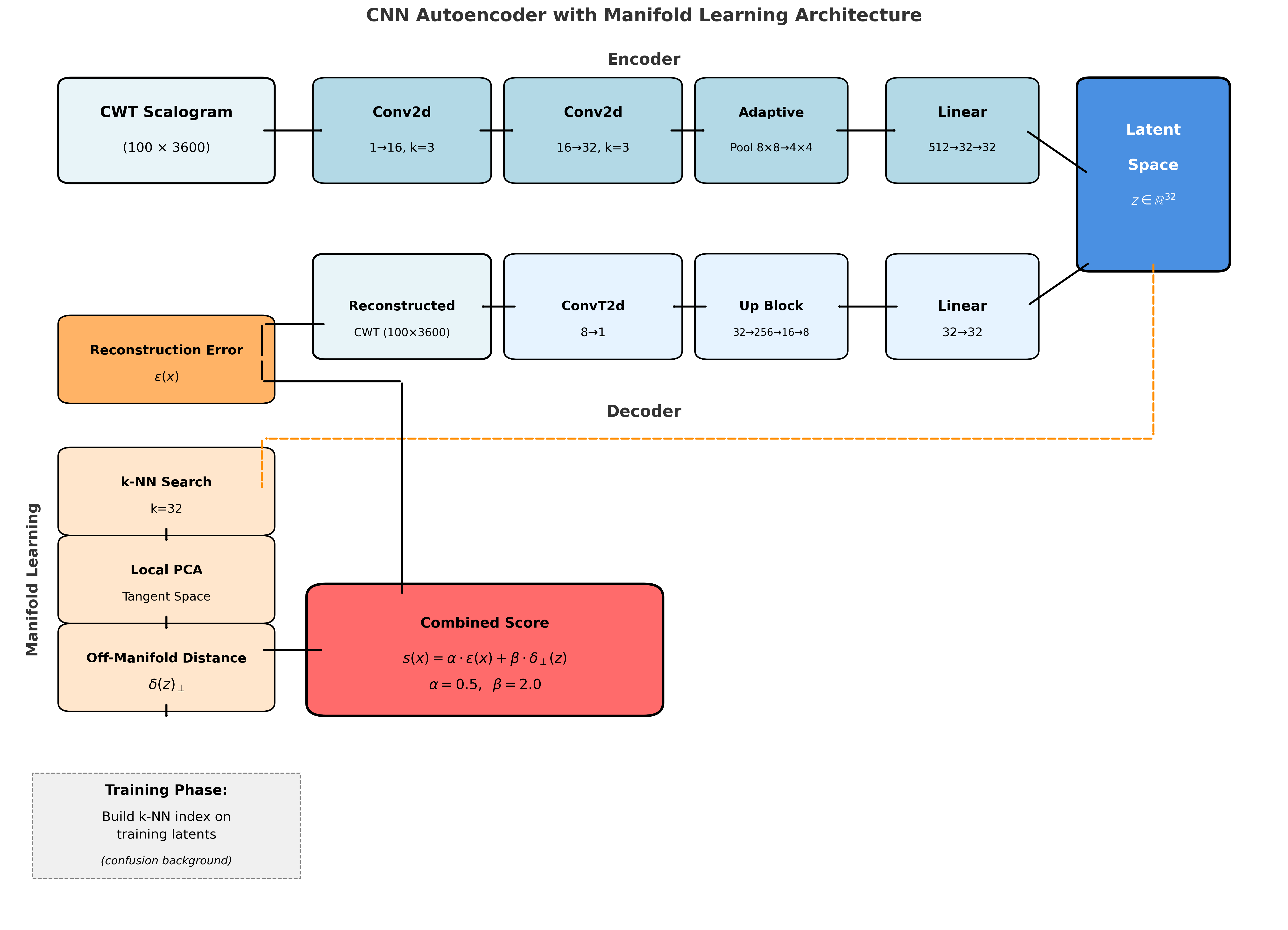}
    \caption[CNN autoencoder architecture with manifold learning]{
        Architecture of the CNN-based autoencoder with manifold learning for LISA
        gravitational-wave source separation. The encoder (top path) processes CWT
        scalograms ($100 \times 3600$) through two convolutional layers (Conv2d) with adaptive
        pooling, followed by linear layers to produce a 32-dimensional latent representation
        $z$. The decoder (bottom path) reconstructs the input through linear and
        transposed convolutional layers. Reconstruction error $\epsilon(x)$ is computed
        from the difference between input and reconstructed scalograms. In parallel, the
        manifold learning branch (left side) operates on the latent space: k-NN search
        (k=32) identifies nearest neighbors from training data, local PCA estimates the
        tangent space, and the off-manifold distance $\delta_{\perp}(z)$ quantifies
        deviation from the learned manifold. The combined anomaly score
        $s(x) = \alpha \cdot \epsilon(x) + \beta \cdot \delta_{\perp}(z)$ with
        $\alpha = 0.5$ and $\beta = 2.0$ (optimal from grid search) leverages both
        reconstruction error and geometric structure to detect resolvable sources in
        confusion background. During training, the k-NN index is built on latents from
        confusion background data only.
    }
    \label{fig:architecture}
\end{figure}

\medskip
The model consists of a convolutional encoder that compresses CWT scalograms to a 32-dimensional latent space, and a transposed convolutional decoder that reconstructs scalograms from latent representations as shown in Fig.~\ref{fig:architecture}.

The encoder processes CWT scalograms through two-dimensional convolutional layers that extract local time-frequency features. The input is processed through two convolutional blocks: the first uses $3 \times 3$ kernels to expand from 1 to 16 channels, followed by adaptive average pooling to $8 \times 8$ spatial dimensions. The second block expands to 32 channels and pools to $4 \times 4$ dimensions. This progressive downsampling through convolutions and pooling reduces the spatial dimensions from $100 \times 3600$ to $4 \times 4$ while increasing the channel depth, capturing multi-scale time-frequency patterns while preserving local geometric relationships.

The flattened spatial features ($32 \times 4 \times 4 = 512$ dimensions) are then passed through a linear layer that reduces to 32 dimensions, producing the latent representation $z \in \mathbb{R}^{32}$. The decoder mirrors this structure using transposed convolutions: a linear layer expands from 32 to 512 dimensions, followed by reshaping and transposed convolutional layers that upsample back to the original $100 \times 3600$ dimensions. The use of transposed convolutions ensures that the upsampling process also respects local structure, maintaining the geometric properties of the learned representation. The final layer uses a $\tanh$ activation to constrain outputs to $[-1, 1]$, matching the normalized CWT input range.

The model contains approximately 33,000 trainable parameters, making it computationally efficient while maintaining sufficient capacity to learn the structure of confusion background. We train the model to minimize mean squared error between input and reconstructed scalograms, using only training background data (confusion noise plus instrumental noise). This ensures the model learns to reconstruct typical LISA-like background patterns, while resolvable sources embedded in the background will produce higher reconstruction errors.

The latent space $\mathcal{Z} = \mathbb{R}^{32}$ learned by this convolutional encoder becomes the domain for our manifold analysis. The image of the training data under the encoder, $\mathcal{M}_{\mathcal{Z}} = \phi(\mathcal{M})$, forms a $D$-dimensional submanifold of $\mathcal{Z}$ where $D \leq 32$. The convolutional structure ensures that this latent manifold preserves the smooth, locally structured geometry of the time-frequency representation, which is essential for the manifold learning approach described in the next section.

\section*{Manifold Learning in Latent Space}
\label{sec:manifold}

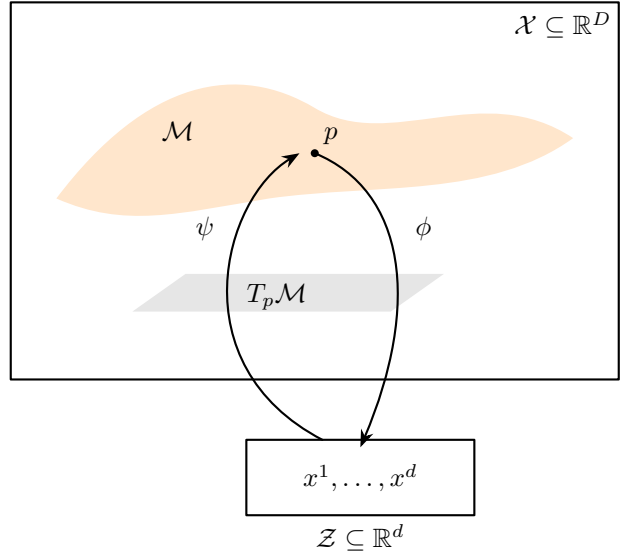
\begin{figure}[htbp]
    \centering
    \begin{tikzpicture}[>=Stealth, line join=round, line cap=round, scale=1.0]

  % Ambient box R^D
  \draw[thick] (0,0) rectangle (8,5);
  \node[anchor=west] at (6.5,4.7) {$\mathcal{X}\subseteq\mathbb{R}^D$};

  % Tangent plane T_pM (slanted quadrilateral)
  \fill[gray!20] (2.3,1.4) -- (5.7,1.4) -- (5.0,0.9) -- (1.6,0.9) -- cycle;
  \node at (3.5,1.1) {$T_p\mathcal{M}$};

  % Manifold M (curved "sheet")
  \begin{scope}
    \fill[orange!20]
      (0.6,2.4)
      .. controls (1.8,4.0) and (3.0,4.2) ..
      (4.0,3.6)
      .. controls (5.0,3.0) and (6.2,4.0) ..
      (7.4,3.2)
      .. controls (6.3,2.4) and (4.8,2.6) ..
      (3.4,2.4)
      .. controls (2.2,2.2) and (1.5,2.0) ..
      (0.6,2.4)
      -- cycle;
  \end{scope}
  \node at (2.2,3.3) {$\mathcal{M}$};

  % Point p on the manifold
  \coordinate (p) at (4.0,3.0);
  \fill (p) circle (1.5pt);
  \node[above right] at (p) {$p$};

  % Coordinate box R^d
  \coordinate (coordboxSW) at (3.1,-0.8);
  \coordinate (coordboxNE) at (6.1,-1.8);
  \draw[thick] (coordboxSW) rectangle (coordboxNE);
  \node at ($(coordboxSW)!0.5!(coordboxNE)$) {$x^1,\dots,x^d$};
  \node[below] at ($(coordboxSW)!0.5!(coordboxNE) + (0,-0.5)$) {$\mathcal{Z}\subseteq\mathbb{R}^d$};

  % Arrow f : M -> R^d
  \draw[thick,->]
    (p)
      .. controls (5.2,2.5) and (5.4,1.0) ..
    ($(coordboxSW)!0.5!(coordboxNE)+(0,0.4)$);
  \node[right] at (5.2,2.0) {$\phi$};

  % Arrow f^{-1} : R^d -> M
  \draw[thick,->]
    ($(coordboxSW)!0.5!(coordboxNE)+ (-0.5,0.5)$)
      .. controls (2.2,0.2) and (2.8,2.4) ..
    (3.8,3.0);
  \node[left] at (2.8,2.0) {$\psi$};

\end{tikzpicture}
    \caption[Embedded manifold and chart]{%
  An embedded manifold $\mathcal{M} \subset \mathcal{X}$ representing the structure
  of high-dimensional data, where $\mathcal{X} \subset \mathbb{R}^D$ is the data space.
  A chart $\phi|_{\mathcal{M}}: \mathcal{M} \to \mathcal{Z}$ maps a point
  $p \in \mathcal{M}$ to low-dimensional coordinates $(x^1,\dots,x^d)$ in
  the latent space $\mathcal{Z} \subset \mathbb{R}^d$, analogous to the encoder
  of an autoencoder. The decoder $\psi: \mathcal{Z} \to \mathcal{M}$ approximates
  the inverse chart $\phi^{-1}$, reconstructing points on $\mathcal{M}$ from their
  latent coordinates. The tangent space $T_p\mathcal{M}$ provides the best linear
  approximation to $\mathcal{M}$ near $p$, highlighting the connection between
  differential geometry and latent-variable models in machine learning.%
  \label{fig:manifold_chart}
}
\end{figure}

\medskip
The autoencoder encoder $\phi$ maps the high-dimensional CWT scalogram space $\mathcal{X}$ 
to a low-dimensional latent space $\mathcal{Z} = \mathbb{R}^{32}$. The choice of latent 
dimension (32) is an architectural hyperparameter of the autoencoder and should not be 
interpreted as the intrinsic dimension of the confusion-background manifold. Under the manifold 
hypothesis, the training data (confusion background) lies near a smooth embedded submanifold 
$\mathcal{M} \subset \mathcal{X}$. The image of this manifold under the encoder, 
$\mathcal{M}_{\mathcal{Z}} = \phi(\mathcal{M}) \subset \mathcal{Z}$, forms a $D$-dimensional 
submanifold where $D \leq 32$ (see Figure~\ref{fig:manifold_chart}). We construct a discrete 
approximation of this latent manifold and use its local differential geometry to augment the 
anomaly score beyond reconstruction error alone.

\subsection*{Manifold Construction}

\medskip
Given training latent vectors $\{z_i = \phi(x_i)\}_{i=1}^N$ where $x_i$ are confusion background segments, we construct a discrete approximation $\mathcal{M}_{\mathcal{Z}}^N$ of the latent manifold using a $k$-nearest neighbors graph. For each training point $z_i$, we identify its $k$ nearest neighbors $\mathcal{N}_k(z_i) = \{z_{i_1}, \ldots, z_{i_k}\}$ using Euclidean distance in $\mathcal{Z}$. This graph structure approximates the manifold as a simplicial complex, where edges connect nearby points that lie on the same local patch of the manifold.

The choice of $k = 32$ neighbors balances local geometry estimation with robustness to noise. Too few neighbors yield unstable tangent space estimates, while too many neighbors blur the local structure. The $k$-NN graph provides a discrete representation of the manifold topology, capturing the connectivity and local structure of the confusion background in latent space.

\subsection*{Local Tangent Space Estimation}

\medskip
At each point $z \in \mathcal{Z}$, we estimate the local tangent space $T_z \mathcal{M}_{\mathcal{Z}}$ using principal component analysis on the $k$ nearest neighbors. This follows from the fundamental theorem of differential geometry: a smooth $D$-dimensional manifold is locally diffeomorphic to $\mathbb{R}^{D}$, and the tangent space at each point is a $D$-dimensional linear subspace of the ambient space.

For a query point $z$ (which may be off-manifold), we first find its $k$ nearest neighbors $\mathcal{N}_k(z) = \{z_{i_1}, \ldots, z_{i_k}\}$ from the training set. We compute the local mean:

\begin{equation}
\mu_z = \frac{1}{k} \sum_{j=1}^k z_{i_j}
\end{equation}

which serves as a reference point on the manifold. The centered neighbors $v_j = z_{i_j} - \mu_z$ for $j = 1, \ldots, k$ approximate tangent vectors at $\mu_z$. The covariance matrix of these centered neighbors is:

\begin{equation}
C = \frac{1}{k-1} \sum_{j=1}^k v_j v_j^T
\end{equation}

Performing principal component analysis on $C$ yields the eigendecomposition $C = U \Lambda U^T$, where $U = [u_1, \ldots, u_d]$ contains the eigenvectors and $\Lambda = \text{diag}(\lambda_1, \ldots, \lambda_d)$ contains the eigenvalues in descending order. The top $D$ eigenvectors, where $D$ is chosen such that $\sum_{i=1}^{D} \lambda_i / \sum_{i=1}^d \lambda_i \geq 0.95$ (or fixed at $D = 8$ in our implementation), span the estimated tangent space:

\begin{equation}
T_{\mu_z} \mathcal{M}_{\mathcal{Z}}^N = \text{span}\{u_1, \ldots, u_{D}\}
\end{equation}

The tangent basis matrix $U_{D} = [u_1, \ldots, u_{D}] \in \mathbb{R}^{d \times D}$ provides an orthonormal basis for the tangent space. The orthogonal complement, spanned by $\{u_{D+1}, \ldots, u_d\}$, is the normal space $N_{\mu_z} \mathcal{M}_{\mathcal{Z}}^N = (T_{\mu_z} \mathcal{M}_{\mathcal{Z}}^N)^{\perp}$.

\subsection*{Geometric Decomposition and Off-Manifold Distance}

\medskip
For any latent point $z \in \mathcal{Z}$, we decompose the displacement vector $r = z - \mu_z$ into tangent and normal components using orthogonal projection. The tangent projection operator is:

\begin{equation}
\Pi_{\parallel}(r) = U_{D} U_{D}^T r
\end{equation}

which projects $r$ onto the tangent space. The normal projection operator is:

\begin{equation}
\Pi_{\perp}(r) = (I - U_{D} U_{D}^T) r = r - \Pi_{\parallel}(r)
\end{equation}

which projects $r$ onto the normal space. This gives the geometric decomposition:

\begin{equation}
z - \mu_z = r_{\parallel} + r_{\perp}
\end{equation}

where $r_{\parallel} = \Pi_{\parallel}(r) \in T_{\mu_z} \mathcal{M}_{\mathcal{Z}}^N$ is the tangent component and $r_{\perp} = \Pi_{\perp}(r) \in N_{\mu_z} \mathcal{M}_{\mathcal{Z}}^N$ is the normal component.

The normal deviation, or off-manifold distance, is defined as:

\begin{equation}
\delta_{\perp}(z) = \|r_{\perp}\| = \|(I - U_{D} U_{D}^T)(z - \mu_z)\|
\end{equation}

This measures the Euclidean distance from $z$ to the tangent space at $\mu_z$, which approximates the distance from $z$ to the manifold itself. Geometrically, $\delta_{\perp}(z)$ quantifies how far the point deviates from the manifold in the normal direction, perpendicular to the local tangent space. Points that lie on the manifold should have $\delta_{\perp}(z) \approx 0$, while points that are off-manifold (such as resolvable sources embedded in confusion background) should have large $\delta_{\perp}(z)$.

\subsection*{Combined Anomaly Score and Coefficient Interpretation}

\medskip
We combine the autoencoder reconstruction error with the geometric off-manifold distance to form a composite anomaly score. For a test point $x \in \mathcal{X}$ with latent representation $z = \phi(x)$, the combined score is:

\begin{equation}
s(x) = \alpha \cdot \epsilon(x) + \beta \cdot \delta_{\perp}(\phi(x))
\end{equation}

where $\epsilon(x) = \|x - \psi(\phi(x))\|^2$ is the reconstruction error and $\delta_{\perp}(z)$ is the off-manifold distance. The manifold learning layer is shown in Fig.~\ref{fig:architecture}.

The coefficient $\alpha$ weights the reconstruction error term. This measures how well the autoencoder can reconstruct the input scalogram through the learned chart $\psi \circ \phi$. Large reconstruction error indicates that the input deviates from the patterns the model learned during training on confusion background. The reconstruction error captures global representability: whether the entire scalogram can be faithfully reconstructed from the latent representation.

The coefficient $\beta$ weights the off-manifold distance term. This measures the geometric deviation of the latent representation from the learned manifold structure. Large $\delta_{\perp}(z)$ indicates that the latent point lies in the normal bundle rather than on the manifold itself, suggesting the point is geometrically distinct from the confusion background. The off-manifold distance captures local geometric structure: whether the latent representation respects the smooth manifold geometry learned from training data.

The relative magnitudes of $\alpha$ and $\beta$ determine the balance between these two complementary sources of information. If $\beta = 0$, the score reduces to standard autoencoder-based anomaly detection using only reconstruction error. If $\beta > 0$, the score incorporates geometric information from the latent manifold, testing whether the manifold hypothesis provides additional discriminative power beyond reconstruction alone. The optimal values of $\alpha$ and $\beta$ are determined through grid search, and the value of $\beta_{\text{opt}}$ directly quantifies the contribution of manifold geometry to anomaly detection performance.

\section{Experimental Setup}
\label{sec:experimental}

We conduct experiments on synthetic LISA data generated as described in Section~\ref{sec:data}. The training set consists of 5000 background segments containing instrumental noise plus confusion noise from 1000 unresolved galactic binaries with SNR in the range $[0.1, 2.0]$. The test set contains 200 background segments and 400 signal segments, where signal segments include a resolvable source (massive black hole binary, extreme mass ratio inspiral, or galactic binary) embedded in confusion background with SNR in the range $[10, 50]$. The signal type distribution is 50\% massive black hole binaries, 30\% extreme mass ratio inspirals, and 20\% galactic binaries. All segments have duration 3600 seconds sampled at 1 Hz.

CWT preprocessing is applied using the parameters described in Section~\ref{sec:preprocessing}: frequency range $[10^{-4}, 10^{-1}]$ Hz, 140 frequency scales (approximately 14 scales per octave), Morlet wavelet, and target dimensions $100 \times 3600$. Global normalization statistics are computed from raw time-domain training data before CWT preprocessing.

The autoencoder architecture (Section~\ref{sec:autoencoder}) processes $100 \times 3600$ CWT scalograms through an encoder-decoder structure with 32-dimensional latent space, containing approximately 33,000 trainable parameters. Training hyperparameters are summarized in Table~\ref{tab:training_params}. The model is trained for 100 epochs with batch size 4, using the Adam optimizer with learning rate $0.001$ and mean squared error loss. A validation split of 20\% is used for early stopping with patience 7 epochs and minimum improvement threshold $0.0005$. The learning rate is reduced by a factor of 0.5 when validation loss plateaus.

\begin{table}[htbp]
\centering
\caption{Training hyperparameters}
\label{tab:training_params}
\begin{tabular}{lc}
\hline
Parameter & Value \\
\hline
Epochs & 100 \\
Batch size & 4 \\
Learning rate & 0.001 \\
Optimizer & Adam \\
Loss function & MSE \\
Dropout & 0.1 \\
Validation split & 0.2 \\
Early stopping patience & 7 epochs \\
Early stopping min $\Delta$ & 0.0005 \\
Learning rate scheduler & ReduceLROnPlateau \\
Weight decay & $10^{-5}$ \\
\hline
\end{tabular}
\end{table}

After training, latent representations are extracted for 4000 training samples (excluding 1000 validation samples). The latent manifold is constructed using $k$-nearest neighbors with $k = 32$ and Euclidean distance metric, as described in Section~\ref{sec:manifold}. Local tangent spaces are estimated using principal component analysis with tangent dimension $D = 8$.

We perform a grid search over the coefficient space $(\alpha, \beta)$ to determine the optimal balance between reconstruction error and off-manifold distance. The search space is $\alpha \in \{0.5, 1.0, 2.0, 5.0, 10.0\}$ and $\beta \in \{0.0, 0.01, 0.05, 0.1, 0.5, 1.0, 2.0\}$, yielding 35 total combinations. For each combination, we compute the combined anomaly score $s(x) = \alpha \cdot \epsilon(x) + \beta \cdot \delta_{\perp}(\phi(x))$ for all test samples and evaluate performance using ROC-AUC. The combination achieving the highest ROC-AUC is selected as optimal. The grid search includes $\beta = 0$ as a baseline corresponding to autoencoder-only detection without manifold geometry.

The implementation uses PyTorch \cite{Paszke2019} for model training, NumPy \cite{Harris2020} and SciPy \cite{Virtanen2020} for numerical computations, and scikit-learn \cite{Pedregosa2011} for nearest-neighbor search and principal component analysis.

\section{Results}
\label{sec:results}

The optimal configuration achieves maximum ROC-AUC at $\alpha = 0.5$ and $\beta = 2.0$, indicating that manifold geometry contributes four times more weight than reconstruction error in the combined anomaly score.

\begin{figure}[htbp]
    \centering
    \includegraphics[width=0.7\textwidth]{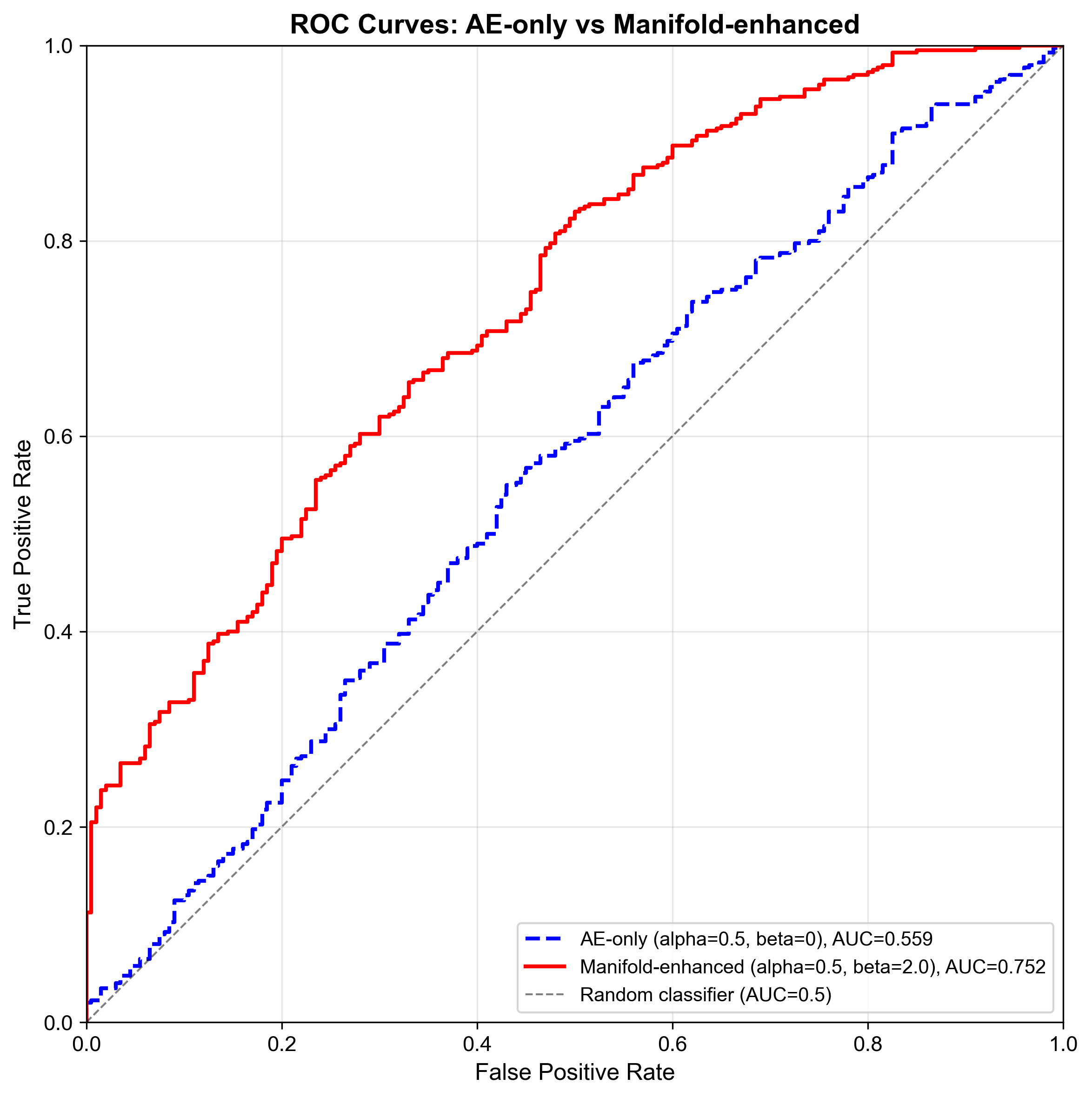}
    \caption[ROC curves: AE-only vs manifold-enhanced]{
        Receiver Operating Characteristic (ROC) curves comparing autoencoder-only
        detection (alpha=0.5, beta=0, dashed blue line, ROC-AUC=0.559) with
        manifold-enhanced detection (alpha=0.5, beta=2.0, solid red line, ROC-AUC=0.752).
        The diagonal dashed line represents a random classifier (ROC-AUC=0.5). The
        manifold-enhanced approach achieves significantly higher true positive rates
        across all false positive rates, demonstrating the benefit of incorporating
        geometric information from the latent space manifold. The improvement in ROC-AUC
        from 0.559 to 0.752 represents a 35\% relative improvement over the
        autoencoder-only baseline.
    }
    \label{fig:roc_curves}
\end{figure}

\begin{figure}[htbp]
    \centering
    \includegraphics[width=0.7\textwidth]{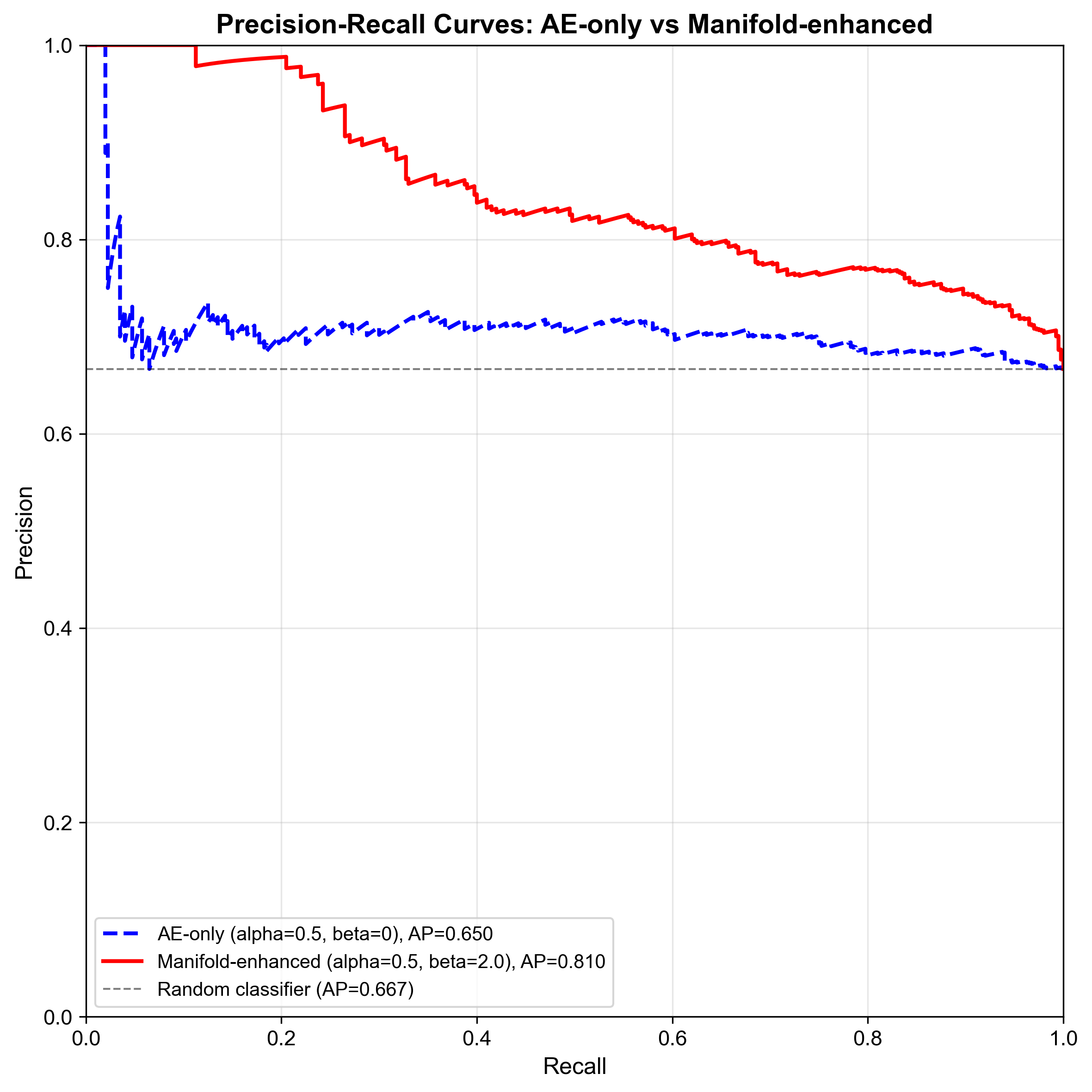}
    \caption[Precision-Recall curves: AE-only vs manifold-enhanced]{
        Precision-Recall curves comparing autoencoder-only detection (alpha=0.5, beta=0,
        dashed blue line) with manifold-enhanced detection (alpha=0.5, beta=2.0,
        solid red line). The horizontal dashed line represents the baseline precision
        of a random classifier. The manifold-enhanced approach maintains higher precision
        across all recall values, indicating better discrimination of resolvable sources
        from confusion background. The area under the precision-recall curve (AP) is
        0.650 for autoencoder-only detection and 0.810 for manifold-enhanced detection.
        At the threshold that optimizes F1 score, the manifold-enhanced method
        achieves precision=0.81 and recall=0.61, compared to precision=0.71 and
        recall=0.54 for the autoencoder-only baseline.
    }
    \label{fig:pr_curves}
\end{figure}

Table~\ref{tab:results} summarizes performance metrics for the optimal configuration and the autoencoder-only baseline. Importantly, both methods use identical data, the same trained autoencoder, and the same reconstruction errors; the only difference is that the combined method augments the autoencoder scores with manifold geometry information. The manifold-enhanced method achieves ROC-AUC = 0.752 and AP = 0.810, while the autoencoder-only baseline ($\beta = 0$) achieves ROC-AUC = 0.559 and AP = 0.650. At the threshold that optimizes F1 score, the manifold-enhanced method achieves precision = 0.81 and recall = 0.61, compared with precision = 0.71 and recall = 0.54 for the autoencoder-only baseline. These results indicate that manifold geometry provides a $35\%$ relative improvement in ROC-AUC and improvements in both precision and recall as shown in Figs.~\ref{fig:roc_curves} and \ref{fig:pr_curves}. This improvement cannot be attributed to stochastic variation or different models, as it results directly from incorporating geometric structure into the scoring function.

% Figure 5: Precision-Recall Curves Comparison
% PLACEMENT: Results section, when discussing precision and recall metrics

\begin{table}[htbp]
\centering
\caption{Performance metrics for optimal configuration ($\alpha = 0.5$, $\beta = 2.0$) and autoencoder-only baseline ($\beta = 0$). *Precision and recall are evaluated at the threshold that optimizes F1 score.}
\label{tab:results}
\begin{tabular}{lcccc}
\hline
Method & ROC-AUC & AP & Precision* & Recall* \\
\hline
AE-only ($\beta = 0$) & $0.559$ & $0.650$ & $0.71$ & $0.54$ \\
AE + Manifold (optimal) & $0.752$ & $0.810$ & $0.81$ & $0.61$ \\
\hline
\end{tabular}
\end{table}

\begin{figure}[htbp]
    \centering

    \begin{subfigure}[b]{0.60\textwidth}
        \centering
        \includegraphics[width=\textwidth]{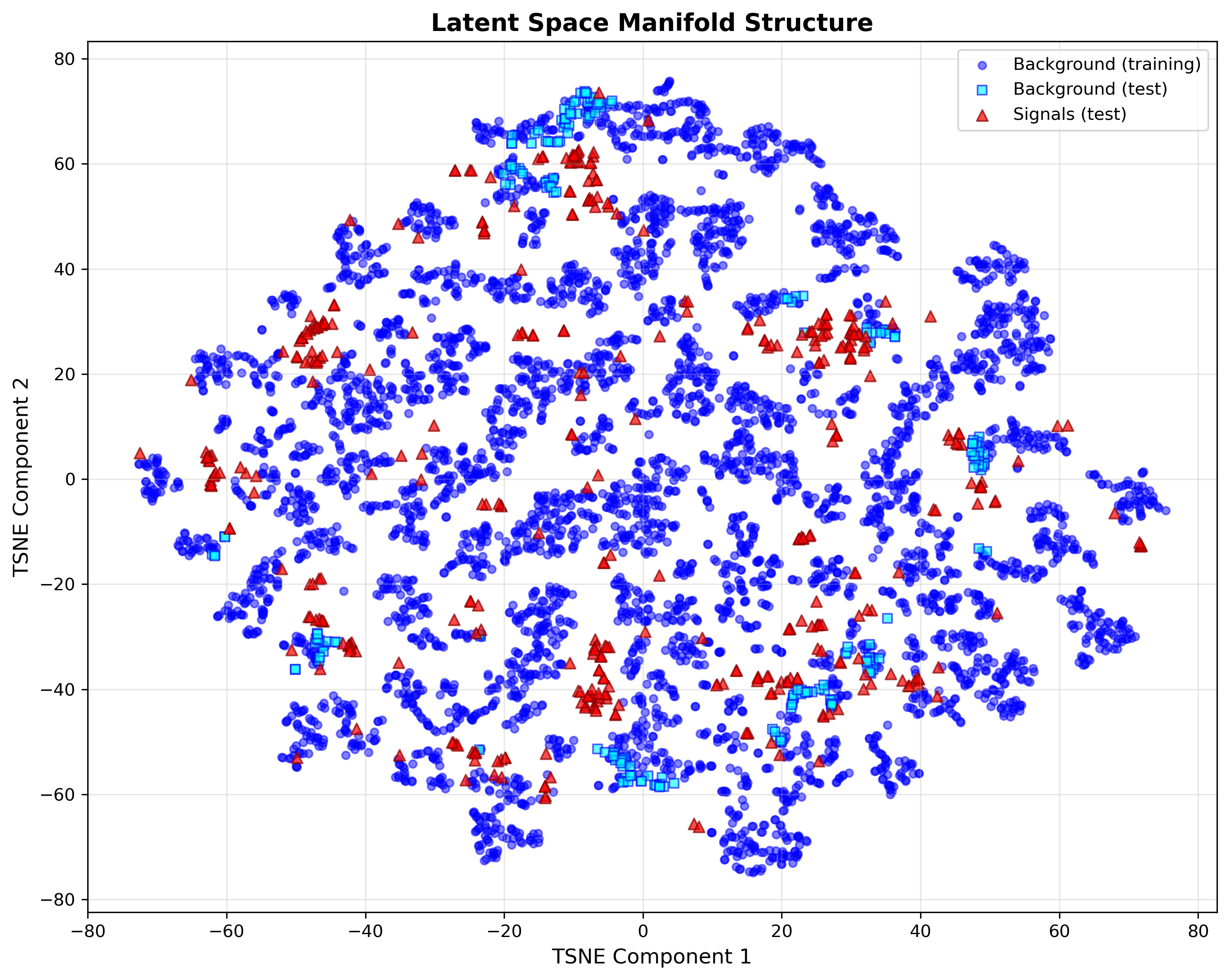}
        \caption{2D t-SNE projection}
        \label{fig:latent_2d}
    \end{subfigure}

    \vspace{0.5em}

    \begin{subfigure}[b]{0.60\textwidth}
        \centering
        \includegraphics[width=\textwidth]{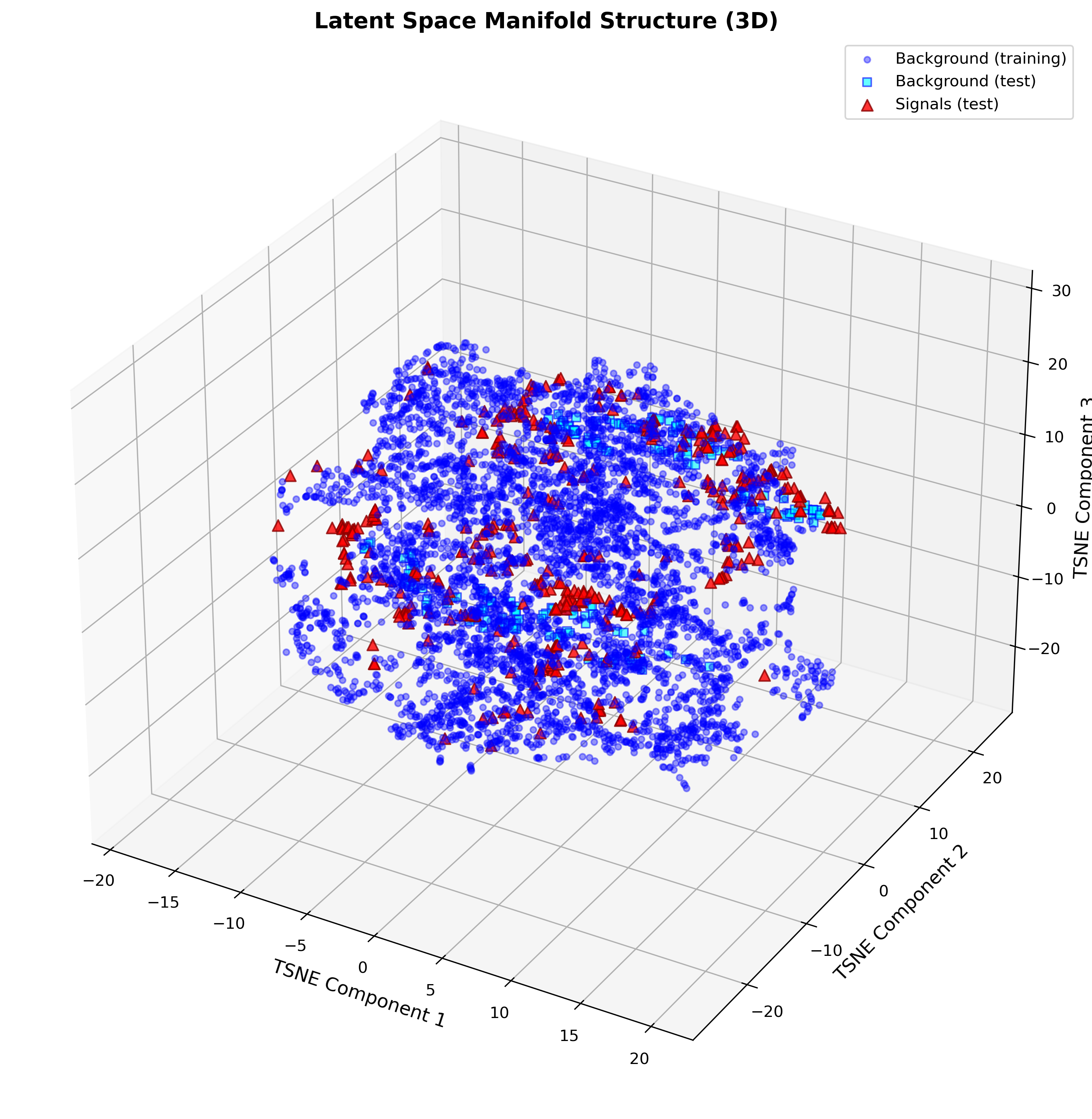}
        \caption{3D t-SNE projection}
        \label{fig:latent_3d}
    \end{subfigure}

    \caption[Latent space manifold structure]{
        t-SNE projections of the 32-dimensional latent space learned by the autoencoder,
        showing the manifold structure of confusion background and resolvable sources.
        Training background samples (blue circles) form a dense, connected manifold
        representing typical LISA confusion noise. Test background samples (cyan squares)
        are embedded within this manifold, confirming that the model generalizes to unseen
        confusion background. Resolvable signals (red triangles) are distributed throughout
        the latent space, with many points lying off the learned manifold. (a) The 2D projection
        shows the overall structure but signals may appear embedded due to projection artifacts.
        (b) The 3D projection reveals that signals form a diffuse halo around the central
        background cluster, with the vast majority located at the surface or periphery rather
        than buried deep within it. This geometric structure---signals at the manifold boundary
        rather than completely isolated---demonstrates that signals are off-manifold but remain
        in proximity to the background distribution, consistent with the source separation
        challenge in LISA's confusion-limited regime. The visualization is based on 5000 training
        samples and 600 test samples (200 background, 400 signals).
    }
    \label{fig:latent_manifold}
\end{figure}

The learned manifold structure is visualized in Figure~\ref{fig:latent_manifold},
which shows a t-SNE projection of the 32-dimensional latent space. Training 
background samples (blue circles) form a dense, connected manifold representing 
typical LISA confusion noise. Test background samples (cyan squares) lie within 
this manifold, confirming that the model generalizes to unseen confusion background. 
Resolvable signals (red triangles) are distributed throughout the latent space, 
with many points lying off the learned manifold, indicating geometric distinction 
from confusion background. This geometric structure, captured by tangent space 
analysis, provides additional discriminative power beyond reconstruction error 
alone, contributing to the improved performance compared to 
autoencoder-only detection.

The optimal coefficients ($\alpha = 0.5$, $\beta = 2.0$) demonstrates that the geometric structure of the latent space provides significant discriminative power for source separation in LISA data. The presence of many overlapping galactic binaries in the confusion background creates a structured manifold in latent space, and resolvable sources (massive black hole binaries, extreme mass ratio inspirals) appear as off-manifold anomalies. The four-fold greater weight on geometric distance compared to reconstruction error indicates that the manifold structure captures information about the typical confusion background that is not fully captured by reconstruction error alone, enabling better separation of resolvable sources from the confusion-limited signal.

\begin{figure}[htbp]
    \centering
    \includegraphics[width=0.8\textwidth]{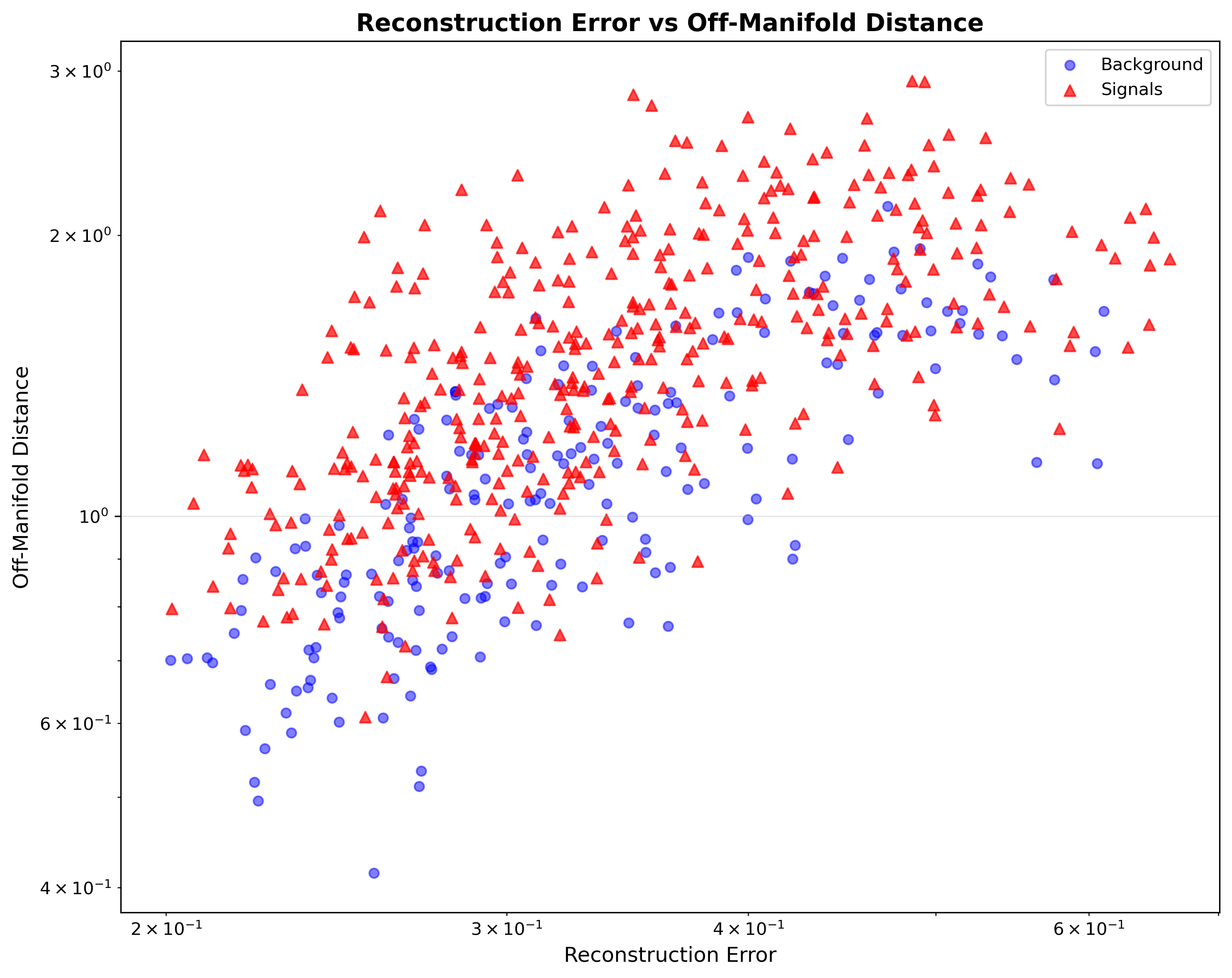}
    \caption[Reconstruction error vs off-manifold distance]{
        Two-dimensional scatter plot showing reconstruction error (x-axis) versus
        off-manifold distance (y-axis) for test samples. Background samples (blue circles)
        cluster in the lower-left region, indicating low reconstruction error and
        small off-manifold distance, consistent with lying on the learned manifold.
        Signal samples (red triangles) are distributed across both dimensions, with
        many samples showing high reconstruction error, high off-manifold distance,
        or both. The optimal combined score $s(x) = \alpha \cdot \epsilon(x) + \beta \cdot \delta_{\perp}(\phi(x))$
        with $\alpha = 0.5$ and $\beta = 2.0$ leverages both dimensions to achieve
        ROC-AUC = 0.752, significantly outperforming autoencoder-only detection (ROC-AUC = 0.559).
        The separation between background and signals in this two-dimensional space
        demonstrates that manifold geometry provides complementary information to
        reconstruction error, enabling better discrimination of resolvable sources
        from confusion background. Both axes are shown on logarithmic scale.
    }
    \label{fig:reconstruction_manifold}
\end{figure}

Figure~\ref{fig:reconstruction_manifold} shows the relationship between 
reconstruction error $\epsilon(x)$ and off-manifold distance $\delta_{\perp}(\phi(x))$ 
for test samples. Background samples (blue circles) cluster in the lower-left 
region, indicating low reconstruction error and small off-manifold distance, 
consistent with lying on the learned manifold. Signal samples (red triangles) 
are distributed across both dimensions, with many samples showing high 
reconstruction error, high off-manifold distance, or both. The optimal combined 
score $s(x) = \alpha \cdot \epsilon(x) + \beta \cdot \delta_{\perp}(\phi(x))$ 
with $\alpha = 0.5$ and $\beta = 2.0$ leverages both dimensions to achieve 
ROC-AUC = 0.752, significantly outperforming autoencoder-only detection (ROC-AUC = 0.559). 
The separation between background and signals in this two-dimensional space 
demonstrates that manifold geometry provides complementary information to 
reconstruction error, enabling better discrimination of resolvable sources 
from confusion background.

% Figure 3: Reconstruction Error vs Off-Manifold Distance

\section{Discussion}
\label{sec:discussion}

The optimal configuration of coefficients ($\alpha = 0.5,\;\beta=2.0$) indicates that incorporating manifold geometry improves discrimination between confusion background and resolvable sources in this dataset. Within the scoring convention used in this study, the optimal configuration assigns a larger coefficient to the off-manifold distance term than to the reconstruction error. This suggests that geometric structure in the latent space carries information about the typical confusion background that is not captured
by reconstruction error alone, indicating that the autoencoder learns a
latent representation in which the confusion background forms a
structured manifold. The relative magnitudes of $\alpha$ and $\beta$ should not be interpreted as
intrinsic physical weights, as their values depend on the normalization and
parameterization of the reconstruction error and manifold distance terms.

The improvement in ROC-AUC from $0.559$ to $0.752$ and in AP from $0.650$ to $0.810$ represents a meaningful enhancement in detection capability considering that both methods use identical data and the same trained autoencoder. At the threshold that optimizes F1 score, precision improves from $0.71$ to $0.81$ while recall improves from $0.54$ to $0.61$, indicating that manifold geometry helps reduce false positives while maintaining or slightly improving true positive detection rates. 

The success of manifold learning in LISA's confusion-limited scenario can be understood through the structure of the data. The confusion background consists of many overlapping galactic binaries with similar frequency characteristics. This creates a structured distribution in the latent space that forms a learnable manifold. Resolvable sources such as massive black hole binaries and extreme mass ratio inspirals have distinct frequency evolution that pushes them off this manifold. 

In addition to its interpretive value, the approach may also have a practical role within larger LISA data-analysis pipelines. Once trained, the model evaluates new segments through a single forward pass of the network and a local geometric calculation in the latent space, making the anomaly score relatively inexpensive to compute. This suggests a potential use as a fast pre-search or anomaly-screening stage that highlights segments of the data stream likely to contain resolvable sources. Such candidates could then be passed to more computationally intensive inference pipelines, such as reversible-jump Markov chain Monte Carlo (RJMCMC) or global Bayesian fit (GBF) methods, which attempt to jointly model large populations of overlapping signals. In this sense the method should be viewed as complementary to existing LISA analysis techniques rather than as a replacement for them.

Several limitations should be acknowledged. The present study is a controlled proof-of-concept designed to test whether latent-space geometry and density structure contain useful information for distinguishing resolvable sources in a confusion-limited setting. The synthetic data generation used here adopts a simplified confusion model with a fixed population of unresolved galactic binaries per segment and a single-channel TDI representation. Realistic LISA observations will instead contain astrophysically motivated source populations with non-uniform distributions, overlapping source classes, evolving foreground structure, and correlated information across the full TDI observables (A,E,T). In addition, the experimental comparison in this work is limited to a baseline convolutional autoencoder framework in order to isolate the contribution of the proposed latent-space scoring method. While this is appropriate for demonstrating the core idea, comparisons against a broader set of anomaly-detection and representation-learning approaches would help clarify the relative strengths and limitations of the method.

The present experiments also remain modest in scale relative to anticipated LISA data volumes. Although follow-on experiments with multiple random seeds produced qualitatively similar performance statistics, suggesting that the observed behavior is not strongly dependent on initialization, larger-scale validation will still be necessary. Future work will therefore focus on extending the framework to more realistic simulation environments, including astrophysically motivated galactic foreground models and community benchmark datasets such as the LISA Data Challenge datasets. Additional directions include multi-channel analysis using the full TDI observables, evaluation against broader baseline methods, scaling to larger training corpora, and investigation of alternative architectures including transformer-based and sequence-aware models. Such studies will help determine whether the latent-space structure identified here persists under realistic LISA observing conditions.
\section{Conclusion}
\label{sec:conclusion}

We have demonstrated that manifold learning provides significant benefit for source separation in confusion-limited gravitational-wave data. The optimal combination with $\beta = 2.0$ achieves ROC-AUC = 0.752 and AP = 0.810, with a $35\%$ improvement in ROC-AUC over autoencoder-only detection. This result establishes that geometric structure in the latent space captures information about the confusion background that aids in identifying resolvable sources. The success of this approach in this confusion-limited scenario suggests that manifold learning may be particularly valuable for source separation problems in gravitational-wave astronomy.

\section{Data and Code Availability}

All code, data processing scripts, trained models, and results presented in this study are publicly available for reproducibility and further research. The complete implementation is archived with a persistent DOI at:

\begin{center}
\url{https://doi.org/10.5281/zenodo.18912193}
\end{center}

The source repository is hosted on GitHub at:

\begin{center}
\url{https://github.com/jericho-cain/cwt-manifold-grav-wav}
\end{center}

The archived release corresponds to the version of the code used to generate the results presented in this paper.

The repository includes:

\begin{itemize}
\item Complete synthetic LISA data generation pipeline with instrumental noise, confusion noise, and waveform generators (MBHB, EMRI, galactic binaries)
\item LISA noise model implementation following standard PSD models
\item Continuous wavelet transform (CWT) preprocessing pipeline adapted for the LISA frequency range ($10^{-4}$ to $10^{-1}$ Hz)
\item CNN-based autoencoder implementation in PyTorch for CWT scalogram reconstruction
\item Manifold learning framework with k-nearest neighbors and tangent space estimation
\item Training and evaluation scripts with comprehensive logging and run management
\item Grid search implementation for $\alpha$ and $\beta$ coefficient optimization
\item Automated testing suite with unit and integration tests
\item Data validation tools for synthetic LISA data quality assurance
\end{itemize}

Synthetic LISA datasets are generated on demand using the included data-generation scripts. This allows the full analysis pipeline to be reproduced without distributing large datasets and ensures transparency in the construction of the simulated confusion-limited data used in this study.

\section{Acknowledgments}

The author thanks the LISA mission team and the gravitational-wave community for their work on LISA science requirements and data analysis frameworks that informed this study. Special appreciation goes to the developers of PyTorch, NumPy, SciPy, scikit-learn, and other open-source libraries that enabled this research.

The author also acknowledges the gravitational-wave community for developing the theoretical foundations and analysis methods that motivated this work, particularly the development of template-free anomaly detection approaches for gravitational-wave data.

\bibliographystyle{iopart-num}
\bibliography{refs}

\end{document}